\documentclass[12pt]{article}
\usepackage{amsmath,amsfonts,amssymb,amsthm}
\usepackage{xy}
\xyoption{all}

\advance \topmargin by -\headheight
\advance \topmargin by -\headsep
\evensidemargin \oddsidemargin
\def\R{\mathbb{R}}
\def\Z{\mathbb{Z}}
\def\C{\mathbb{C}}

\def\A{\mathcal{A}}
\def\G{\mathcal{G}}
\def\B{\mathcal{B}}

\def\Lie{\operatorname{Lie}}
\def\Ad{\operatorname{Ad}}
\def\GL{\operatorname{GL}}
\def\im{\operatorname{im}}
\def\End{\operatorname{End}}
\def\Cov{\operatorname{Cov}}
\def\Homeo{\operatorname{Homeo}}

\def\o{\otimes}

\def\sp{{\operatorname{Spin}}}
\def\spv{{\operatorname{Spin}(V)}}
\def\spc{{\operatorname{Spin}^c}}
\def\spcv{{\operatorname{Spin}^c(V)}}
\def\spn{{\operatorname{Spin}(n)}}
\def\spcn{{\operatorname{Spin}^c(n)}}

\def\clrn{\operatorname{C\ell}(\R^n)}
\def\clorn{\operatorname{C\ell}^0(\R^n)}

\def\clo{{\operatorname{C\ell}^0}}
\def\cloe{{\operatorname{C\ell}^0(E)}}
\def\cl{{\operatorname{C\ell}}}
\def\dslash{{\not\!\partial}} %Clifford differential operator (Feynman slash \partial)

\def\pso{P_{SO}}
\def\psp{P_{\sp}}

\def\Prin{\operatorname{Prin}}
\def\Tr{\operatorname{Tr}}

\def\g{\mathfrak{g}}

\def\eps{\epsilon}

\def\Aut{\operatorname{Aut}}
\def\d{\partial}
\def\del{\nabla}
\def\ad{\operatorname{ad}}
\def\aut{\operatorname{aut}}
\def\vol{\operatorname{vol}}
\def\tri{\triangle}

\def\To{\longrightarrow}

\def\te#1{\text{#1}}
\def\norm#1{\left\| #1 \right\|}
\def\abs#1{\left| #1 \right|}

\newtheorem{theorem}{Theorem}

\theoremstyle{definition}
\newtheorem{definition}{Definition}
\theoremstyle{remark}

\begin{document}

\title{Gauge Theory: Instantons, Monopoles, and Moduli Spaces}
\author{William Gordon Ritter \\ Jefferson Physical Laboratory \\ Harvard University, Cambridge, MA}
\date{\today}
\maketitle

\begin{abstract}
In this expository review we discuss various aspects of gauge
theory. While the focus is on mathematics, wherever possible we
make contact with theoretical high energy physics. Particular
emphasis is placed on instantons and monopoles, which admit
physical interpretation, and yield interesting and nontrivial
mathematics. We give a clear and essentially self-contained
exposition of the mathematical structure of the Seiberg-Witten
monopole equations. Other topics include Donaldson's theorem on
moduli spaces of monopoles, compactification of spaces of ASD
connections, The Abelian monopole equations, and Abelian Higgs
vortices.
\end{abstract}

\section{Introduction}

In this section we introduce the basic notation and setup for the later sections,
which will discuss instantons and monopoles both from the viewpoint of physics,
and from the viewpoint of relevance to Floer homology.

\subsection{Gauge Theory}
The mathematical scenario in which we work is that of Yang-Mills-Higgs theory,
which in $d$ space dimensions is defined by a gauge potential
\[
A = A_j(x) \, dx^j
\]
and a scalar Higgs field
\[
\Phi = \Phi(x)
\]
The components $A_j(x)$ take values in the Lie algebra $\g$ of a finite dimensional
Lie group $G$, the gauge group. The field $\Phi$ takes values in a
representation space $L$ of $G$, corresponding to a representation $\rho : G \to \Aut(L)$.
We regard the book \cite{JT} as a fundamental reference for the
global analysis of these structures.

The gauge potential defines a curvature
\[
F = dA + A \wedge A = \frac{1}{2} F_{ij}(x) \, dx^i \wedge dx^j
\]
where in the last expression, the components $F_{ij}$ are given
by the equations
\[
F_{ij}(x) = \d_i A_j(x) - \d_j A_i(x) + [A_i(x), A_j(x)]
\]
The connection (Lie algebra-valued 1-form) $A$ defines a covariant
derivative acting on the Higgs field, by the following equation
\[
D_A\Phi = (\del_A)_j (\Phi) dx^j
\]
where $(\del_A)_j (\Phi) = \del_j \Phi + \rho(A_j)(\Phi)$.
This is a special case of the general possibility of defining
an exterior covariant derivative on $p$-forms taking values in a
representation of a Lie algebra ($\g$-module). For an $L$-valued
$p$-form $\omega$, we define
\[
D_A\omega \equiv d\omega + \rho(A) \wedge \omega
\]
This is covariant with respect to the natural action of a gauge
transformation $g : \R^d \to G$ in the following way:
\[
D_{A_g}(\rho(g)\omega) = \rho(g) \cdot D_A\omega
\]
With these definitions, the Euclidean Yang-Mills-Higgs action is
\begin{equation} \label{YMH-action}
\A(A, \Phi) = \frac{1}{2} \int_{\R^d} \left\{
(F_A, F_A) + (D_A\Phi, D_A\Phi) + \frac{\lambda}{4} \ \ast (\abs{\Phi}^2-1)^2 \right\}
\end{equation}
where $\lambda \geq 0$ is a coupling constant and the last term in
\eqref{YMH-action} represents the self-interaction of the Higgs field.

In more mathematical language, $A$ is a connection in a principal
$G$-bundle $P$ over a manifold $M$. $\Phi$ is a section of the
adjoint bundle $\ad(P) = P\otimes_{\ad} \g$. The gauge
transformations $g$ are taken as smooth sections of the bundle
$\aut(P)$. In a fundamental paper \cite{F1}, Floer studies actions
of the form \eqref{YMH-action} with Higgs coupling $\lambda = 0$.

The question is to characterize the field configurations $c$ which
minimize the action functional
\begin{equation} \label{action}
\A(c) = \int_M (\abs{F_A}^2 + \abs{d_A\Phi}^2) \, d\vol_M
\end{equation}
It is well known that for $M = \R^3$, the absolute minima of
\eqref{action} are characterized by the Bogomolny equations
\begin{equation} \label{bogomolny}
d_A \Phi = \pm \, \ast F_A
\end{equation}
An interesting aspect of Yang-Mills-Higgs theory is that the
moduli space of solutions of \eqref{bogomolny}, modulo gauge
equivalence, is a finite dimensional manifold for a large class
of interesting examples.

With the insertion of the Higgs self-interaction term
$\frac{\lambda}{4} \ \ast (\abs{\Phi}^2-1)^2$, the variational
equations for \eqref{YMH-action} are given by
\begin{eqnarray}
D_A \, \ast F &=& \ast J \label{var-1} \\
\del_A^2 \Phi &=& \frac{\lambda}{2} \Phi(\abs{\Phi}^2 - 1) \label{var-2}
\end{eqnarray}
where $J$ is known in the physics literature as the \emph{current}, and
is defined for an arbitrary representation $\rho$ of the Gauge group
by the equation
\begin{equation} \label{current}
J = -(\rho(h^a) \Phi, \, D_A \Phi) \, h_a
\end{equation}
in which $\{h^a : a = 1, \ldots, \dim\g \} = \{h_a\}$
is a chosen basis of $\g$. The solutions to equations \eqref{var-1}, \eqref{var-2},
\eqref{current} do not depend on this choice.

\subsection{Instantons}
\label{sec:instantons} In this section we begin our discussion of
instantons. Recall that finite-action solutions of the field
equations \eqref{var-1}, \eqref{var-2}, \eqref{current} are called
\emph{solitons}. In the physics literature, the term
\emph{instanton} refers to a soliton in $d=4$ Euclidean ``pure
Yang-Mills theory'' (that is, the theory that results from
\eqref{YMH-action} by setting $\lambda = 0$ and $\Phi \equiv 0$,
in other words the variational theory of connections on a
principal bundle).

The following one-instanton solution of pure Yang-Mills theory was
given by Belavin, Polyakov, Tyupkin, and Schwartz. Consider the
gauge group $G = SU(2)$ and let $g : \R^4 \to SU(2)$ be given by
$g = \abs{x}^{-1} (x^0 + i x^k \sigma_k)$, where $\sigma^1,
\sigma^2, \sigma^3$ are the Pauli matrices. Then $A =
\frac{x^2}{x^2 + \mu^2} g \, dg^{-1}$ is a finite-action field
configuration satisfying the pure Yang-Mills equations. This
solution was found by explicitly computing the curvature two-form
\begin{equation} \label{instanton-curv}
F = \sum_{\ell = 1}^3 \tau_\ell \left( \frac{2}{r}
\frac{df}{dr} e^4 \wedge e^\ell + \frac{2}{r^2} (f^2 - f)
\sum_{j,k=1}^3 \eps_{jk\ell} e^j \wedge e^k \right)
\end{equation}
with $\tau_j = -\frac{1}{2} \sigma_j$, where $\sigma_j$ are the
Pauli matrices.  One then explicitly computes the Hodge star of
\eqref{instanton-curv} and shows that the self-duality equation $F
= \ast F$ leads to the following differential equation for $f$:
\[
\frac{df}{dr} = -\frac{2}{r} f(f-1)
\]
which has solution $f(r) = \frac{r^2}{r^2 + c^2}$, $c \in \R$.

\subsection{The Instanton Bundle}
In Section \ref{sec:instantons} above, we introduced the instanton
potential,
\[
A_1 = \frac{r^2}{r^2 + c^2} \gamma^{-1} d\gamma
\]
as a solution to the Euclidean Yang-Mills equations. The map
$\gamma : \R^4 \setminus \{0\} \To SU(2)$ is given by
\[
\gamma(x) = \frac{1}{r} \Big( x^4 - i \sum_{j=1}^3 \sigma_j x^j \Big),
\]
and $\sigma_j$ are the Pauli matrices. The potential $A_1$ is
regular at $x = 0$, but decays only as $O(r^{-1})$ at infinity.
However, we can change these features with a gauge transformation.
Explicitly,
\[
A_2 = \gamma A_1 \gamma^{-1}  + \gamma d\gamma^{-1} =
\frac{c^2}{r^2 + c^2} \gamma d\gamma^{-1}
\]
and note that this $A_2$ is now singular at $x = 0$, but
vanishes as $O(r^{-3})$ at infinity.
Let $U_1 = S^4 - \{\te{south pole}\}$ and
$U_2 = S^4 - \{\te{north pole}\}$ denote the standard covering of
$S^4$ by two charts, where each chart is identified with $\R^4$
by means of stereographic projection.
We can thus consider $A_j (j=1,2)$ as being defined on $U_j$
via pullback. The slow decay of $A_1$ at $r \to \infty$
means that $A_1$ cannot be extended across the south pole.
Similarly, the singularity of $A_2$ at the origin prevents
its being extended from $U_2$ to all of $S^4$ (but the rapid
decrease at infinity implies that its stereographic projection
is well defined across the north pole). It now follows as a
special case of a general existence theorem for connections on
principal bundles
(Theorem \ref{thm:prin-charts}, reproduced below) that $A_1$ and $A_2$
are local representatives of a connection on an $SU(2)$
bundle over $S^4$ whose transition function is $\gamma$.
The total space of this bundle is $S^7$.

\begin{theorem} \label{thm:prin-charts}
Assume $P \overset{\pi}{\To} M$ is a principal $G$-bundle.
Let $\{U_r\}$ be an open covering of $M$. Given a family of local
$\g$-valued 1-forms $A_r \in \Lambda^1(U_r, \g)$ which fulfill the
compatibility condition:
\[
\text{ for } x \in U_r \cap U_s, \quad
A_{r, x} = \Ad(g_{sr}^{-1}(x)) A_{s, x} + (g^*_{sr} \zeta)_x
\]
where $\zeta$ is the Maurer-Cartan form on $G$, and given a set
of local sections $\sigma_r : U_r \to \pi^{-1}(U_r)$ satisfying
\[
\sigma_s(x) = \sigma_r(x) g_{rs}(x)
\]
there is a unique connection $\A$ on $P$ such that $A_r = \sigma_r^* \A$.
\end{theorem}

\subsection{Monopoles}
In this section we begin our discussion of monopoles.
We again consider a special case of the Yang-Mills-Higgs variational equations,
in which now $d = 3$, $G = U(1)$, the representation is the adjoint representation
$L = \g$, $\rho(x) = \ad(x)$ and $\lambda = 0$, so the Higgs field self-interaction
is infinitely weak but the gauge potential still couples to the
Higgs field through the covariant derivative terms.

This is the mathematical model of the physical phenomena of electromagnetism.
In this model, $A$ corresponds to the ``magnetic vector potential'' encountered
in Maxewell theory, and $\Phi$ is the scalar electric potential.

Introduce a density $q_e(x)$ of electric charge, and $q_m(x)$ a corresponding
magnetic charge density. In the presence of these background charges the
static Maxwell equations are given by
\begin{eqnarray} \label{maxwell-1}
d \, \ast F &=& 0 \\
dF &=& (4\pi) \ast q_m  \label{maxwell-2} \\
-\tri \Phi &=& 4\pi q_e  \label{maxwell-3}
\end{eqnarray}
One obtains Dirac's magnetic monopole by considering the case in
which $q_m(x) = \delta(x - x_0)$.  This allows for the possibility
that the integral of the closed 2-form $F$ over a two-sphere in
$\R^3 \setminus \{x_0\}$ is nonzero, which gives in turn a nonzero
divergence of the magnetic field (which is, by definition $\vec B
= \ast F = \ast dA$).

All $U(1)$ monopole field configurations have infinite action; no
finite action solutions to \eqref{maxwell-1}-\eqref{maxwell-3}
exist when $q_m(x)$ is a delta function.

\subsection{The Monopole Bundle}
In order to describe a magnetic monopole within the framework of
electrodynamics, Dirac used the vector potential
\[
A = i \frac{m}{2r(x^3-r)} (x^1 dx^2 - x^2 dx^1)
\]
in which $x^0, \ldots, x^4$ are coordinates in Minkowski
space, $r^2 = (x^1)^2 + (x^2)^2 + (x^3)^2$ denotes the
squared distance to the origin in Euclidean space, and $m$
is an integer constant.

Introducing polar coordinates $(r, \vartheta, \varphi)$ on space,
and noting that in the polar expression for $A$ and its gauge
transformations, $r$ does not appear. Thus we can consider $A$ (and
gauge transformations $A' = A + \gamma^{-1} d\gamma$ as living
on the 2-sphere. Indeed, on $U_1 = S^2-\{(0,0,-1)\}$ we have
\[
A_1 = \frac{1}{2} i \, m \, (1-\cos\vartheta) d\varphi
\]
and similarly on
$U_2 = S^2-\{(0,0,1)\}$ we have
\[
A_2 = \frac{1}{2} i \, m \, (-1-\cos\vartheta) d\varphi
\]
On the overlap $U_1 \cap U_2$ the connections are related by
a gauge transformation $A_1 = A_2 + \gamma^{-1} d\gamma$,
where $\gamma$ is some function on spacetime taking
values in $U(1)$. An appeal to Theorem \ref{thm:prin-charts}
shows that $A_1$ and $A_2$ determine a connection on a
$U(1)$ principal bundle over $S^2$ whose transition function
$g_{21} : U_1 \cap U_2 \to U(1)$ is given by $\gamma$.
If $m \ne 0$, the bundle is nontrivial, and in general, different
values of $m \in \Z$ do not give rise to equivalent bundles.

We give more details in the case $m = 1$. The total space of
the bundle is $S^3$, and identifying $S^3$ with the unit sphere
in $\C^2$ gives coordinates
\[
z^1 = y^1 + i y^2, \quad z^2 = y^3 + i y^4
\]
which can be used to define a (Lie-algebra valued) connection 1-form
\[
\A := i (y^1 dy^2 - y^2 dy^1 + y^3 dy^4 - y^4 dy^3)
\]
which takes values in the Lie algebra $u(1)$. By direct
calculation one verifies that the local representatives of
$\A$ are the potentials $A_1$ and $A_2$ of the $m=1$
Dirac monopole.

\section{Bundles and Covariant Derivatives}

A principal $G$-bundle $P$ over a smooth manifold $X$ is a
manifold with a smooth right $G$-action, and with
$X = $ the orbit space $P/G$. The action must be locally equivalent
to the obvious action on $U \times G$ where $U$ is an open set in $X$.
This local product structure defines a fibration $\pi : P \to X$.

We will describe here three equivalent ways to describe a connection
on such a bundle:

\begin{enumerate}
\item[(i)]  A smooth distribution of $G$-invariant `horizontal subspaces'
$H \subset TP$, where `horizontal'
means that for all $p \in P$, we have a decomposition
$TP_p = H_p \oplus T(\pi^{-1}(x))$, where $x = \pi(p)$.

\item[(ii)]  A $G$-invariant 1-form $A$ on $P$ taking values in the Lie algebra $\g = \Lie(G)$,
or in other words a section of the bundle $T^*P \otimes \g$, where `$G$-invariant'
refers to a combination of the given action of $G$ on $P$, and the
adjoint action representation of $G$ on $\g$.

\item[(iii)]  A covariant derivative $\del$ on an associated vector bundle $E$,
which is by definition a linear map $\del : \Omega_X^0(E) \To \Omega_X^1(E)$
satisfying the Leibniz rule: $\del(f \, s) = f\del s + (df) \, s$ where
$f$ is a function on $X$, and $s \in \Gamma(E)$. This gives a connection
in the sense of (i), as follows. Let $P$ be the frame bundle of $E$, then
a local section $\sigma$ of $P$, which is a collection $(s_1, \ldots, s_n)$
of local sections of $E$, is called `horizontal' if $(\del s_i)_x = 0$
for all $i = 1, \ldots, n$. Then define $H_p$ to be the tangent space to
(the image of) a horizontal
section $\sigma$ through $p$, regarding the latter as a submanifold of $P$.
\end{enumerate}

For the classical groups, these are all equivalent. In working with the
Yang-Mills equations on Euclidean spaces, the most useful connection
is the standard product connection which can be defined on any
trivial bundle $\C^n \times X$ by taking the covariant derivative
to be the usual notion of `total derivative' of vector valued functions.

If $X$ is an oriented Riemannian four-manifold, Hodge theory gives
a decomposition of the 2-forms on $X$ into self-dual and
anti-self-dual (ASD) pieces, which are just the $\pm 1$ eigenspaces
of the Hodge $\ast$ operator:
\[
\Omega_X^2 = \Omega_X^+ \oplus \Omega_X^-
\]
This construction extends naturally to bundle-valued forms, and
hence to the curvature form $F_A$ of a connection.
The connection is called ASD (self-dual) if $F_A^+ (F_A^-)$ is zero.

For connections with $SU(r)$ structure group on a 4-manifold $X$, we have
\[
c_2(E) = \frac{1}{8\pi^2} \int_X \Tr(F_A^2) \ \in \ \Z
\]
and a connection is ASD if and only if we have
$\Tr(F_A^2) = \abs{F_A}^2 \, d\mu$, where $d\mu$ is the Riemannian
volume element. We now make connection with Yang-Mills theory:
the functional
\[
\norm{F_A}^2 = \int_A \Tr(F_A^2) \, d\mu
\]
is a special case of the Yang-Mills functional, and these observations
suffice to show that $\abs{8\pi^2 c_2(E)}$ gives a lower bound on the
Yang-Mills functional, which (when $c_2 > 0$) is achieved precisely
in case that the connection $A$ is anti-self-dual.

\section{Spin, Spin${}^{\mathbf c}$, and Dirac}

The material in this section owes much to \cite{LM}.

Let $E \to X$ denote a real $n$-dimensional vector bundle
over the smooth manifold $X$.
We assume:
\begin{enumerate}
\item  We have a positive definite inner product $(\ ,\ )$
defined continuously in the fibers.
\item  The bundle is \emph{oriented}, so there is a choice
of orientation for the vector space which forms each fiber, chosen
in a continuous way.
\end{enumerate}
We will investigate the second condition more closely.
Note that in general, we have an isomorphism
\begin{equation} \label{H1-iso}
H^1(X; G) \cong \left\{
\begin{matrix}
\te{ equivalence classes of } \\
\te{ principal $G$-bundles on $X$ }
\end{matrix}
\right\}
\end{equation}
This implies that the set $\Cov_2(X)$ of equivalence classes of
2-sheeted coverings of $X$ is isomorphic to $H^1(X; \Z_2)$.
Let $P_{O} \to X$ be the principal $O(n)$ bundle whose fiber
at $x \in X$ is the space of ON bases of $E_x$.
The bundle of orientations $O_E$ is the quotient
$P_O/SO(n)$, which is a 2-sheeted covering of $X$.
In fact, $E \to X$ is orientable iff $w_1(E) = 0$,
where $w_1$ denotes the $1^{st}$ Stiefel-Whitney class,
which is the element of $H^1(X; \Z_2)$ defined by
the covering $O_E \to X$ via the isomorphism \eqref{H1-iso}.

Thus choosing an orientation is equivalent to
simplifying the structure group of the bundle from $O(n)$ to its subgroup
$SO(n)$, which is connected. We now ask whether it is possible
to obtain a simply connected structure group.
Let $P_{SO} \to X$ denote the bundle of \emph{oriented} orthonormal
frames in $E$, and for $n \geq 3$ let $\xi_0 : \spn \to SO(n)$
denote the universal covering group homomorphism.

\begin{definition}
A \emph{spin structure} on $E$ is a principal $\spn$ bundle
$P_\sp(E)$ together with a 2-sheeted covering
$\xi : P_\sp(E) \to P_{SO}(E)$ such that
\[
\xi(pg) = \xi(p) \xi_0(g) \quad \te{ for all } p \in P_\sp(E), g \in \spn
\]
\end{definition}

Spin structures on $\xymatrix@1{E \ar[r]^{\pi} & X}$
are in natural 1-1 correspondence with the 2-sheeted
coverings of $P_{SO}(E)$ which are non-trivial on fibers of $\pi$.
Also, there is an existence-uniqueness result for spin structures
similar to that which was noted above for orientations.
A spin structure exists $\iff$ the $2^{nd}$ Stiefel-Whitney
class $w_2(E)$ is zero, and in this situation the distinct
spin structures lie in 1-1 correspondence with the elements
of $H^1(X; \Z_2)$. In summary, if a Riemannian vector
bundle $E$ over $X$ is equivalent to a vector bundle with
connected, simply connected structure group, then it is
orientable and spin. A manifold $X$ is said to be \emph{spin}
if $TX \to X$ is a spin bundle.

We now discuss briefly the associated bundle construction.
Given a principal $G$-bundle $P \to X$, $G$ a lie group,
and given a continuous homomorphism $\rho : G \To \Homeo(F)$
from $G$ into the group of homeomorphisms of a space $F$, then
consider the free left action of $G$ on $P \times F$ given
by
\begin{equation} \label{eqn:assoc-bndle}
g \cdot (p, f) := (p g^{-1}, \rho(g)f)
\end{equation}
The projection $\xymatrix@1{P \times F \ar[r] & P \ar[r]^{\pi} & X}$
is constant on orbits under \eqref{eqn:assoc-bndle}, hence it
induces a mapping $\xymatrix@1{P \times_\rho F \ar[r]^{\pi_\rho}  & X}$
where $P \times_\rho F$ is the orbit space of the action.
If $\rho : G \To \GL(V)$ is a linear representation, then
$(P \times_\rho F, \pi_\rho)$ is a vector bundle.

Now let $M$ be a left module over the Clifford Algebra
$\clrn$, and let $\xi : P_\sp(E) \to P_{SO}(E)$ be a
spin structure. A \emph{real spinor bundle} for $E$
is a bundle of the form $S(E) = P_\sp(E) \times_\mu M$
where $\mu : \spn \to SO(M)$ is the
representation given by left multiplication by elements of
$\spn \subset \clorn$. Similarly, one can consider
complex spinor bundles by considering
$M_\C = $ a complex representation for $\clrn \o \C$.

Let $n = 2m$ be even and let $S_\C(E) = $ the irreducible
complex spinor bundle of $E$. Then $S_\C(E)$ splits naturally
into a direct sum
\[
S_\C(E) = S_\C^+(E) \oplus S_\C^-(E)
\]
of $\cloe$-modules, which are defined to be the $\pm 1$
eigenspaces of Clifford multiplication by the element
defined at each point $x \in X$ by the following equation:
\[
\omega_{\C, x} := i^m e_1 \cdot \ldots \cdot e_{2m}
\]
where $e_1, \ldots, e_{2m}$ is a positively oriented basis of $E_x$.

We now remark on the important special case $E = TX$.
We set $\pso(X) = \pso(TX)$,  $\cl(X) = \cl(TX)$.
Then there exists a unique connection on $\pso(X)$ with
identically vanishing torsion.
If $X$ has a spin structure $\xi : \psp(X) \to \pso(X)$
then we can lift this canonical Riemannian connection
to a connection on $\psp(X)$. Thus all spin bundles
inherit this connection.

In general, let $S$ be any bundle of left modules over
$\cl(X)$. Assume $S$ is Riemannian and has the canonical
connection. In this scenario, there is a canonical first
order differential operator called the \emph{Dirac operator}
\[
D : \Gamma(S) \To \Gamma(S)
\]
by using the following local formula: at $x \in X$ we define
\[
D\sigma := e^j \cdot \nabla_{e_j} \sigma
\]
where $\{e^i\}$ is an ON basis of $TX_x$ and $\cdot$ denotes
Clifford multiplication. By direct computation of symbols, one checks that
both $D$ and $D^2$ (the \emph{Dirac laplacian}) are elliptic operators.
The Dirac operator in this form was first written down by Atiyah and
Singer in the course of their work on the index theorem.

Finally, we remark that one can develop much of the theory of
spin structures in the parallel case of $\spc$ structures, in which
(roughly speaking) many of the important structural elements are
complexified. By definition the group $\spcv$ is the subgroup
of the multiplicative group of units of $\cl(V) \o_\R \C$
generated by $\spv$ and the unit circle of complex scalars.
Then there is an isomorphism $\spcv \cong \sp(V) \times_{\Z_2} U(1)$.
[Proof: since the circle of unit scalars commutes with $\spv$, it
follows that we have a natural surjective map $\sp(V) \times U(1) \To \spcv$.
The kernel of this map is $\{(\alpha, \alpha^{-1}) \mid \alpha \in \sp(V) \cap U(1) \}$,
but the intersection of $\sp(V)$ with the scalars is $\{\pm 1\} = \Z_2$. ]

Thus, $\spcv$ is the double covering group of $SO(V) \times U(1)$
which is nontrivial on each factor.
\begin{definition}
Let $\pso \to X$ be a principal $SO(n)$-bundle on $X$.
A \emph{$\spc$ structure} on $\pso$ consists of a principal
$U(1)$ bundle $P_{U(1)}$ and a principal $\spcn$-bundle
$P_\spc$ together with a $\spc$-equivariant bundle map
\[
P_\spc \To \pso \times P_{U(1)}
\]
The integral class $c \in H^2(X; \Z)$ corresponding
to $P_{U(1)}$ under the isomorphism
$H^2(X; \Z) \cong \Prin_{U(1)}(X)$ is called the
\emph{canonical class} of the $\spc$ structure.
\end{definition}

\section{Moduli Spaces of Monopoles: Donaldson's Theorem}

A monopole on $\R^3$ is defined by Atiyah \cite{A} to consist
of a gauge field (connection) $A_\mu(x), \mu = 1,2,3$, and
a Higgs field $\psi(x)$ (all smooth functions of
$x \in \R^3$ and take values in the Lie algebra of $SU(2)$)
which satisfy the so-called \emph{Bogomolny equations}
\[
D\phi = \ast F
\]
where $D_\mu \phi = \d_\mu\phi + [A_\mu,\phi]$ and
$F_{\mu\nu} = \d_\mu A_\nu - \d_\nu A_\mu + [A_\mu, A_\nu]$.
Moreover, we require the energy ($L^2$-norm of $F$) to be finite.
The Bogomolny equations are in fact equivalent to the
self-dual Yang-Mills equations in Euclidean 4-space.
The general structure of the moduli space of solutions to
these equations was given by Donaldson in \cite{Don}, in which
he gives an identification of a circle bundle over the moduli
space with the space of basepoint preserving rational maps
$\C P^1 \to \C P^1$ of degree $k$. To mention a few of the
basic details which are used in Donaldson's proof fits well
into this survey. The starting point is a theorem of Hitchin
\cite{Hit}
\begin{theorem}[Donaldson]
There is a natural equivalence between the following two structures:
\begin{enumerate}
\item  monopoles for the group $SU(2)$ with charge $k$, up to
gauge transformation.

\item conjugacy classes under $O(k, \R)$ of matrix-valued functions
$f_1(s), f_2(s), f_3(s)$ of one real variable $s \in (0,2)$ satisfying
\begin{eqnarray*}
\frac{dT_i}{ds} + \sum_{j,k} \eps_{ijk} T_j T_k &=& 0  \\
T_i^*(s) &=& -T_i(s) \\
T_i(2-s) &=& T_i(s)^t
\end{eqnarray*}
In addition, we require that $T_i$ will extend to a meromorphic
function on a neighborhood of $[0,2]$ with simple poles at
$s = 0, 2$ but otherwise analytic, and the residues of the matrices
$T_i$ at the poles $s=0,2$ define an irreducible representation of $SU(2)$.
\end{enumerate}
\end{theorem}

Donaldson has observed that the conditions
$\frac{dT_i}{ds} + \sum {\eps_{ijk}} T_j T_k = 0$ are  equivalent to the
ASD equations for the connection
\[
\A = T_1(s) dx_1 + T_2(x)dx_2 + T_3(s) dx_3
\]
on $\R \times \R^3$ with coordinates $s, x_1, x_2, x_3$.

\section{Compactification of Moduli Spaces of ASD Connections}

We wish now to introduce moduli spaces of ASD Yang-Mills connections.
Let $E$ be a bundle over a compact oriented Riemannian $X^4$.
The set $M_E$ is defined to be the set of gauge equivalence classes of
ASD connections on $E$. More explicitly, `gauge equivalence' refers to
the action of the bundle automorphism group $\G$.

We recall briefly how the gauge group acts on the space of connections.
For a given trivialization $\tau$ of $E$, we let $A^\tau$ denote the
connection matrices of $A$ in this trivialization.
(By ``connection matrices'' we will always be implicitly referring
to the fact that the connection can be viewed as a one-form
taking values in a finite-dimensional semisimple Lie algebra, i.e. a matrix
Lie algebra.)

Suppose $u \in \G$, the group of bundle automorphisms of $E$.
We define the action of $u$ on the connection $A$ by how it
transforms the corresponding covariant derivatives:
\[
\del_{u(A)} s = u \del_A(u^{-1} s)
\]
Using the fact that $u$ is a section of the vector bundle $\End(E)$,
we can take the covariant derivative of it. In this language,
we have $u(A) = A - (\del_A u) \, u^{-1}$.

In terms of the connection matrices in some trivialization we have
\[
A^{u\tau} = u A^\tau u^{-1} - (du) u^{-1}
\]

The space $\A$ of connections has the structure of an affine space, and
the topological and metric structure of an infinite-dimensional
Banach manifold, however the moduli space $M_E$ with its inherited topology,
is finite dimensional

The construction of this moduli space can be broken naturally into two steps.
\begin{enumerate}
\item[(i)] Find the solutions to the ASD equation. $F_A^+ = 0$,
where the $+,-$ indices refer to the natural splitting of
$\Omega_X^2$ into $\pm 1$ eigenspaces of the Hodge $\ast$ operator.

\item[(ii)] Quotient by the action of the gauge group.
\end{enumerate}

We will now discuss the usefulness of Sobolev spaces in Part (ii)
of this program.

In four dimensions, the Sobolev embedding theorem states that
$k > 2 \implies W^{k,2}$ embeds into the space of continuous functions.
One can then define the notion of a $W^{k,2}$ map from a domain
in $X$ to the structure group $G$ of the bundle.
In many important examples, $G$ can be identified with a group of
unitary matrices, and then one can consider matrix-valued maps
which are of class $L^2_k$ with respect to the norm on $U(n)$.
One then defines an $L^2_k G$-bundle to be a bundle in which the
transition functions are of class $L^2_k$. In a similar line of
thought, one defines Sobolev spaces of connections on a $G$-bundle
by demanding that in any local trivialization the connections are given
by $L^2_{k-1}$ connection matrices. Such connections have curvature
in $L^2_{k-2}$, Here we have used the fact that
pointwise multiplication induces a continuous map
\[
L^2_{k-1} \times L^2_{k-1} \To L^2_{k-2}
\]
for $k > 2$.

We write $\A$ for the space of $L^2_{k-1}$ connections on an
$L^2_{k} G$-bundle, and $\B$ for the quotient $\A/\G$,
where of course $\G$ denotes automorphisms in the category of
$L^2_k$ bundles.

The important point is that the $L^2$ metric on $\A$ is preserved
by the action of $\G$ so that
\[
d([A], [B]) := \inf_{g \in \G} \norm{A - g(B)}
\]
gives a metric on $\B$. In particular
$\B$ is Hausdorff in the $L^2_{k-1}$ topology, which is finer than the
$L^2$ topology.

We now comment briefly on the local structure of the moduli space $\B$.
In light of the topological isomorphism
\[
\Omega^1(\g) = \im d_A \oplus \ker d_A^*
\]
that is given by elliptic theory, it holds that
a neighborhood of $[A]$ in $\B$
can be described as a quotient of $T_{A,\eps}$ for small $\eps$,
where
\[
T_{A,\eps} = \{ a \in \Omega^1(\g) \mid d_A^* = 0, \ \
\norm{a}_{L^2_{k-1}} < \eps \}
\]

For $X$ a compact orientable manifold with Riemannian metric $g$,
and $SU(2)$-bundle $E$ with second Chern class $c_2(E) = k$. If
$b^+ > 0$, the moduli space $M_k$ of ASD connections on $E$ is,
for a generic metric, an orientable smooth manifold of dimension
$8k - 3(1-b_1+b^+)$, where $b^+$ is the number of positive
eigenvalues of the intersection form and $b_1$ is the first Betti
number. The manifold $M_k$ is not necessarily compact, but it can
be compactified (as shown by Uhlenbeck) Further information on
these compactifications can be obtained from \cite{DK} or from
\cite{OGP}.

As a final remark on the moduli space for the ASD equations,
we note that it is not empty.
A theorem of Clifford Taubes \cite{Taubes1} establishes the existence of
self-dual connections on a 4-manifold $M$ whose intersection
form is positive definite, using analytic techniques to build
the connections on $M$ from those on $S^4$. Atiyah, Hitchin, and Singer
were able also to construct these connections, under the additional
assumption that $M$ is ``half-conformally flat,'' using twistor theory
to convert Yang-Mills into a problem in algebraic geometry.

\section{The Abelian Monopole Equations and Seiberg-Witten Theory}

Let $(X,g)$ be a closed oriented Riemannian four-manifold.
Choose a $\spc$ structure on $X$, $c \in \spc(X)$,  let
$L = L_c$ be the corresponding Hermitian line bundle, and let
$S_L^\pm$ denote the corresponding spinor bundles.

A \emph{field configuration} in this setup is a pair
$(A, \psi)$ where $A$ is a unitary connection on $L$ and
$\psi$ is a smooth section of $S_L^+$. The \emph{Seiberg-Witten
equations}, which one should think of as abelian monopole equations,
are differential equations of these field configurations:
\begin{eqnarray}
\dslash_A \psi &=& 0  \label{SW-1} \\
F_A^+ &=& q(\psi) = \psi \o \psi^* - \frac{\abs{\psi}^2}{2} I \label{SW-2}
\end{eqnarray}
The remainder of this section will be devoted to the structure of
these equations. $\dslash_A$ is the Dirac operator associated to
the Levi-Civit\`a connection on the frame bundle of the tangent
bundle, and the connection $A$ on the determinant line bundle of
the $\spc$ structure. The first equation \eqref{SW-1} just says
that $\psi \in \ker \dslash_A$. For the first equation, note that
$S_L^+$ has a hermitian metric, hence we can identify this bundle
with its dual via an anti-hermitian isomorphism. So $\psi^*$
denotes the image of $\psi$ under this isomorphism. Thus,
\[
\psi \o \psi^* \in S_L^+ \o (S_L^+)^* = \End_\C \Big( S_L^+ \Big)
\]
Now recall that for a positive definite real oriented inner
product space $V$, with oriented ON basis
$\{e_1, \ldots, e_n\}$, we define
\[
\omega_\C = i^{[\frac{n+1}{2}]} e_1 \cdot \ldots \cdot e_n
\]
This element $\omega_\C$ squares to 1 and does not depend
on the choice of basis. Moreover, its $\pm 1$ eigenspaces
define a canonical splitting of the complexified Clifford
algebra $\cl(V) \o \C$. We write these eigenspaces
as $(\cl(V) \o \C)^\pm$. Then
Clifford multiplication induces an isomorphism
\begin{equation} \label{cliff-iso-1}
(\clo(P) \o \C)^+ \cong \End_\C \Big( S_L^+ \Big),
\end{equation}
and it is readily seen that
\begin{equation} \label{cliff-iso-2}
\textstyle
(\clo(P) \o \C)^+ \cong \C \left( \frac{1 + \omega_\C}{2} \right)
\oplus (\bigwedge^2_+(TX) \o \C)
\end{equation}
Under the isomorphism $\xymatrix@1{ \C \left( \frac{1 + \omega_\C}{2} \right)
\oplus (\bigwedge^2_+(TX) \o \C) \ar[r] &  \End_\C \Big( S_L^+ \Big) }$
implied by eqns.~\eqref{cliff-iso-1}-\eqref{cliff-iso-2},
$(1 + \omega_\C)/2$ acts
as the identity and the traceless endomorphisms of $S_L^+$
come from elements of $\bigwedge^2_+(TX) \o \C$.
The trace of $\psi \o \psi^*$ is $\abs{\psi}^2$, so
$q(\psi) = \psi \o \psi^* - \frac{\abs{\psi}^2}{2} I$
is traceless and can therefore be identified with a section of
$\bigwedge^2_+(TX) \o \C$. Using the metric to identify
the tangent bundle $TX$ with the cotangent bundle $T^*X$,
$q(\psi)$ can be viewed as a complex valued self-dual 2-form.
The other Seiberg-Witten equation \eqref{SW-2} just says that
$q(\psi)$ is the self-dual part of the curvature form.

An excellent reference for further study is \cite{Morgan}.

\section{The Stability of Magnetic, or Abelian Higgs Vortices}

In this section, we discuss results on the stability of magnetic
(or Abelian Higgs) vortices. These are certain critical points of
the energy functional
\begin{equation}
\label{eq:ac} E(\psi,A) = \frac{1}{2} \int_{{\bf R}^2} \left\{
|\nabla_A \psi|^2
                  + (\nabla \times A)^2 + \frac{\lambda}{4}
                  (|\psi|^2-1)^2 \right\}
\end{equation}
for the fields
\[
A : {\bf R}^2 \rightarrow {\bf R}^2 \;\;\;\;\;
\mbox{  and  }  \;\;\;\;\; \psi : {\bf R}^2 \rightarrow {\bf C}.
\]
Here $\nabla_A = \nabla - iA$ is the covariant gradient, and
$\lambda > 0$ is a coupling constant.  For a vector, $A$, $\nabla
\times A$ is the scalar $\partial_1 A_2 - \partial_2 A_1$, and for
a scalar $\xi$, $\nabla \times \xi$ is the vector $(-\partial_2
\xi, \partial_1 \xi)$. Critical points of $E(\psi,A)$ satisfy the
{\em Ginzburg-Landau} (GL) equations
\begin{equation}
\label{eq:eq1}
  -\Delta_A\psi +
  \frac{\lambda}{2}(|\psi|^2-1)\psi = 0
\end{equation}
\begin{equation}
\label{eq:eq2}
  \nabla \times \nabla \times A
  - \Im(\bar{\psi} \nabla_A \psi) = 0
\end{equation}
where $\Delta_A = \nabla_A \cdot \nabla_A$.

Physically, the functional $E(\psi,A)$ gives the difference in
free energy between the superconducting and normal states near the
transition temperature in the Ginzburg-Landau theory. $A$ is the
vector potential ($\nabla \times A$ is the induced magnetic
field), and $\psi$ is an {\em order parameter}.  The modulus of
$\psi$ is interpreted as describing the local density of
superconducting Cooper pairs of electrons.

The functional $E(\psi,A)$ also gives the energy of a static
configuration in the Yang-Mills-Higgs classical gauge theory on
${\bf R}^2$, with abelian gauge group $U(1)$.  In this case $A$ is
a connection on the principal $U(1)$- bundle ${\bf R}^2 \times
U(1)$, and $\psi$ is the {\em Higgs field} (see \cite{JT} for
details).

A central feature of the functional $E(\psi,A)$ (and the GL
equations) is its infinite-dimensional symmetry group.
Specifically, $E(\psi,A)$ is invariant under $U(1)$ {\em gauge
transformations},
\begin{equation}
\label{eq:g1}
  \psi \mapsto e^{i\gamma}\psi
\end{equation}
\begin{equation}
\label{eq:g2}
  A \mapsto A + \nabla \gamma
\end{equation}
for any smooth $\gamma : {\bf R}^2 \rightarrow {\bf R}$. In
addition, $E(\psi,A)$ is invariant under coordinate translations,
and under the coordinate rotation transformation
\begin{equation}
\label{eq:rot}
  \psi(x) \mapsto \psi(g^{-1}x)  \;\;\;\;\;\;\;\;\;\;
  A(x) \mapsto gA(g^{-1}x)
\end{equation}
for $g \in SO(2)$.

Finite energy field configurations satisfy
\begin{equation}
\label{eq:bc}
  |\psi| \rightarrow 1  \;\;\;\;\; \mbox{ as }
  \;\;\;\;\; |x| \rightarrow \infty
\end{equation}
which leads to the definition of the {\em topological degree},
$\mbox{deg}(\psi)$, of such a configuration:
\[
  \mbox{deg}(\psi) = \mbox{deg} \left(
  \left. \frac{\psi}{|\psi|} \right|_{|x| = R} :
  {\bf S}^1 \rightarrow {\bf S}^1  \right)
\]
($R$ sufficiently large).  The degree is related to the phenomenon
of flux quantization.  Indeed, an application of Stokes' theorem
shows that a finite-energy configuration satisfies
\[
  \mbox{deg}(\psi) = \frac{1}{2\pi} \int_{{\bf R}^2} (\nabla \times A).
\]

We study, in particular, ``radially-symmetric'' or ``equivariant''
fields of the form
\begin{equation}
  \label{eq:psiA}
  \psi^{(n)}(x) = f_n(r)e^{in\theta}  \;\;\;\;\;\;\;\;\;\;
  A^{(n)}(x) = n\frac{a_n(r)}{r} \hat{x}^{\perp}
\end{equation}
where $(r,\theta)$ are polar coordinates on ${\bf R}^2$,
$\hat{x}^{\perp} = \frac{1}{r} (-x_2,x_1)^t$, $n$ is an integer,
and
\[
  f_n, a_n : [0,\infty) \rightarrow {\bf R}.
\]
It is easily checked that such configurations, if they
satisfy~(\ref{eq:bc}), have degree $n$. The existence of critical
points of this form is well-known; they are called {\em
$n$-vortices}.

The main results cited in this section concern the stability of
these $n$-vortex solutions.  Let
\[
  L^{(n)} = \mbox{ Hess } E (\psi^{(n)}, A^{(n)})
\]
be the linearized operator for GL around the $n$-vortex, acting on
the space
\[
  X = L^2({\bf R}^2,{\bf C}) \oplus L^2({\bf R}^2,{\bf R}^2).
\]
The symmetry group of $E(\psi,A)$ gives rise to an
infinite-dimensional subspace of $\ker(L^{(n)}) \subset X$, which
we denote here by $Z_{sym}$. We say the $n$-vortex is (linearly)
{\em stable} if for some $c > 0$,
\[
  L^{(n)}|_{Z_{sym}^{\perp}} \geq c,
\]
and {\em unstable} if $L^{(n)}$ has a negative eigenvalue. A basic
result due to Gustafson and Sigal is the following linearized
stability statement:
\begin{theorem}
\label{thm:main} \begin{enumerate}
\item (Stability of fundamental vortices) \\
For all $\lambda > 0$, the $\pm 1$-vortex is stable.
\item (Stability/instability of higher-degree vortices) \\
For $|n| \geq 2$, the $n$-vortex is
\[
  \left\{  \begin{array}{cc}
  \mbox{ stable }  & \mbox{ for } \lambda < 1  \\
  \mbox{ unstable }  & \mbox{ for } \lambda > 1.
  \end{array} \right.
\]
\end{enumerate}
\end{theorem}

Theorem~\ref{thm:main} is the basic ingredient in a proof of the
nonlinear dynamical stability/instability of the $n$-vortex for
certain dynamical versions of the GL equations. These include the
GL gradient flow equations, the Abelian Higgs (Lorentz-invariant)
equations, and the Maxwell equations coupled to a nonlinear
Schr\"odinger equation.

\begin{theorem}
\label{thm:dyn} As a solution of either (GF) or (AH), the
$n$-vortex is stable (in the sense of Lyapunov) for $\lambda > 1$
or for $\lambda < 1$ and $n = \pm 1$, and unstable for $\lambda >
1$ and $|n| \geq 2$.
\end{theorem}

The statement of theorem~\ref{thm:main} was conjectured
in~\cite{JT} on the basis of numerical observations
(see~\cite{jr}). Bogomolnyi (\cite{bog}) gave an argument for
instability of vortices for $\lambda > 1$, $|n| \geq 2$.

The solutions of (\ref{eq:eq1}-\ref{eq:eq2}) are well-understood
in the case of {\em critical coupling}, $\lambda = 1$. In this
case, the {\em Bogomolnyi method} (\cite{bog}) gives a pair of
first-order equations whose solutions are global minimizers of
$E(\psi,A)$ among fields of fixed degree (and hence solutions of
the the GL equations).  Taubes (\cite{t1,t2}) has shown that all
solutions of GL with $\lambda = 1$ are solutions of these
first-order equations, and that for a given degree $n$, the
gauge-inequivalent solutions form a $2|n|$-parameter family. The
$2|n|$ parameters describe the locations of the zeros of the
scalar field.  This is discussed in more detail in \cite{JT}. We
remark that for $\lambda = 1$, an $n$-vortex
solution~(\ref{eq:psiA}) corresponds to the case when all $|n|$
zeros of the scalar field lie at the origin.

It is observed numerically (eg \cite{jr}) that in fields
containing localized vortices of like-signed winding number, the
vortices attract each other when $\lambda < 1$ (bringing vortex
centres closer together lowers the energy) and repel each other
when $\lambda > 1$ (separating vortex centres lowers the energy).
When $\lambda = 1$ there is no interaction (which allows for the
existence of the stable multi-vortex solutions of Taubes,
described above).

\end{document}